# MODELAGEM MATEMÁTICA E ESTABILIDADE DE SISTEMAS PREDADOR-PRESA

Paulo Laerte Natti

Universidade Estadual de Londrina, Departamento Matemática, Londrina, PR.

Neyva Maria Lopes Romeiro

Universidade Estadual de Londrina, Departamento Matemática, Londrina, PR.

Eliandro Rodrigues Cirilo

Universidade Estadual de Londrina, Departamento Matemática, Londrina, PR.

Érica Regina Takano Natti

Pontifícia Universidade Católica do Paraná, Campus Londrina, Londrina, PR.

Camila Fogaça de Oliveira

Senai – Londrina, Faculdade de Tecnologia, Londrina, PR.

Altair Santos de Oliveira Sobrinho

Universidade Estadual de Londrina, Departamento Matemática, Londrina, PR.

Carolina Massae Kita

Universidade Estadual de Londrina, Departamento Computação, Londrina, PR.

**RESUMO:** Neste trabalho investigou-se a estabilidade e o comportamento assintótico de alguns modelos do tipo Lotka-Volterra. Para isso foi utilizado o método de Liapunov, que consiste em analisar a estabilidade de sistemas de equações diferenciais ordinárias (EDO's) em torno da situação de equilíbrio, quando submetidos a perturbações nas condições iniciais.

**ABSTRACT:** This work investigated the stability and the asymptotic behavior of some Lotka-Volterra type models. We used the Liapunov method which consists in analyzing the stability of systems of ordinary differential equations (ODE's) around the equilibrium, when they submitted to perturbations in the initial conditions.

# 1 INTRODUÇÃO

Os primeiros ensaios sobre a formulação de modelos matemáticos para descrever a dinâmica de populações são datados dos séculos XVIII e XIX. Em 1798, Thomas Robert Malthus (1766-1834) publicou a obra *An Essay on the principle of population* (MALTHUS, 1789), na qual assumia que a variação do crescimento de uma população era proporcional à população em cada instante, o que significava dizer que a população aumentava em crescimento exponencial no decorrer do tempo. Este modelo não previa uma limitação para o crescimento de populações. A previsão da população de acordo com o modelo malthusiano estabelecia números elevados em curtos intervalos de tempo (DE OLIVEIRA, 2011; BASSANEZI, 2010).

Convencido de que o modelo de crescimento de Malthus não era adequado para explicar a expansão demográfica de um país, Pierre-François Verhulst (1804-1849) elaborou considerações complementares às enunciações propostas por Malthus. Verhulst supunha que uma população não poderia crescer indefinidamente, pois existiam inibições naturais em seu crescimento, tais como guerras, epidemias, falta de alimentos, entre outros fatores. Segundo Verhulst, o crescimento populacional tem necessariamente um limite constante, conforme o tempo cresce (VERHULST, 1838; VERHULST, 1847).

No mesmo período foram propostos outros modelos para o crescimento de populações, como o modelo de Gompertz em 1825 (BASSANEZI, 2010), no entanto todos esses modelos descreviam a dinâmica de populações isoladas, ou de populações e espécies que não interagiam. Normalmente populações competem pela sobrevivência.

Define-se competição como uma interação de indivíduos da mesma espécie ou de espécies diferentes, animal ou vegetal. Tanto a competição entre indivíduos de mesma espécie, quanto à de espécies diferentes, favorecem um processo de seleção, que atinge seu ápice, geralmente, com a preservação dos seres mais bem adaptados ao ambiente e com a extinção de indivíduos com baixo poder de adaptação. Em outras situações, o sistema de espécies interagentes pode atingir um equilíbrio, estacionário ou dinâmico, onde as espécies coexistem.

Alfred Lotka (1880-1949) e Vito Volterra (1860-1940) propuseram em 1925 e 1926, respectivamente e individualmente, um modelo para a interação entre espécies. O modelo de equações diferenciais do matemático Vito Volterra pretendia descrever o observado aumento da população de uma espécie de peixe predador, e consequente diminuição da população uma espécie de peixe presa, no Mar Adriático durante a Primeira Guerra Mundial (VOLTERRA, 1926). Simultaneamente, o químico e matemático Alfred Lotka desenvolveu um modelo para descrever reações químicas, nas quais as concentrações dos elementos químicos oscilavam, um processo semelhante àquele que ocorre com populações em competição (LOTKA, 1925). Estes modelos, posteriormente denominados modelos de Lotka-Volterra, serviram de base para os modelos matemáticos posteriores utilizados para descrever a dinâmica de sistemas do tipo predador-presa.

Atualmente, modelos do tipo predador-presa são usados em várias áreas do conhecimento, tais como em ciências biológicas e agrárias no controle biológico de pragas (RAFIKOV; BALTHAZAR, 2005); em ciências econômicas para descrever as flutuações/oscilações em bolsas de valores (LOUZOUN; SOLOMON, 2001; CAETANO; YONEYAMA, 2007) e no estudo de competições de mercados (MORRIS; PRATT, 2003; SPROTT, 2004); em ciências ambientais para descrever, por exemplo, a captura e emissão de carbono (TREVISAN; LUZ, 2007); entre outras aplicações. A utilização desses modelos permite uma avaliação qualitativa e quantitativa do impacto da competição em diferentes populações sejam elas populações de átomos ou moléculas, neurônios, bactérias, pragas ou indivíduos infectados ou grupos econômicos.

No contexto de dinâmica de populações citam-se como exemplos clássicos duas situações: (i) aquela de espécies em competição, onde duas populações interagem competindo por um suprimento comum, geralmente comida, e (ii) aquela em que uma das espécies é predadora da outra espécie, a presa, que se alimenta de outro tipo de comida. O modelo de Lotka-Volterra, ou equações predador-presa, descrevem sistemas do segundo tipo (BOYCE; DIPRIMA, 2006; ODUM; BARRET, 2007).

Uma crítica às equações de Lotka-Volterra é que, na ausência de predadores, a população de presas aumenta sem limites. Este problema pode ser corrigido ao se considerar o efeito natural inibidor que o ambiente tem, devido as suas limitações, sobre uma população crescente. Matematicamente, este efeito inibidor pode ser modelado por meio de um termo do tipo Verhulst para a saturação da população de presas (BOYCE; DIPRIMA, 2006; ODUM; BARRET, 2007).

Outra modificação relevante introduzida no modelo de Lotka-Volterra foi o efeito da resposta dos predadores às mudanças na população de presas, efeito chamado de resposta funcional. Assim, modelos de Lotka-Volterra com diferentes tipos de respostas funcionais puderam descrever particularidades encontradas em sistemas predador-presa específicos. Vários cientistas desenvolveram tais modelos, dentre eles citamos: Gause (1934), Holling (1965), Rosenzweig, MacArthur (1963), Tanner (1975), entre outros. Uma revisão sobre esses modelos com diversas respostas funcionais é encontrada nas referências (HASTINGS, 1997; MAY, 2001).

Outra importante contribuição para o entendimento da interação entre predadores e presas foram os estudos experimentais desenvolvidos pelo biólogo Carl Barton Huffaker (1958). Ele observou que em habitats homogêneos os predadores extinguem rapidamente a população de presas, de modo que a heterogeneidade dos habitats é fundamental para a coexistência das espécies por longo período de tempo. Em particular, a heterogeneidade espacial das populações torna-se importante quando o tamanho da população de presas torna-se pequeno ou quando populações perdem contato espacial por algum período de tempo.

Derivados dos estudos de Huffaker foram desenvolvidos os modelos de metapopulações (HANSKI, 1997) e de *patches* (LEVIN; POWEL, 1993). Uma metapopulação consiste de um grupo de subpopulações, ou populações locais, da mesma espécie, separadas espacialmente, que interagem em algum nível. Uma porção ou fragmento (*patch*) de habitat é qualquer área que é usada por uma espécie para reprodução ou obtenção de recursos. Entre tais porções de habitat, embora separadas espacialmente, podem ocorrer migrações, uma vez que as condições e recursos não se encontram homogeneamente distribuídos ao longo da paisagem (habitat). Se os membros individuais das espécies podem se mover entre *patches*, isso é benéfico para a sobrevivência da metapopulação, pois permite a recolonização dos *patches* onde tenha ocorrido uma extinção da população local. O desenvolvimento da teoria de metapopulações enfatizou a importância da conectividade entre populações aparentemente isoladas.

Nas metapopulações as extinções são recorrentes. A estocasticidade ambiental (variações ambientais imprevisíveis), aliada a processos emigratórios, podem levar à extinção em uma dada porção (*patch*) do habitat. Por outro lado, processos migratórios amortecem os efeitos de variações ambientais estocásticas. Atualmente, modelos de metapopulações com dinâmica estocástica local têm sido largamente utilizados para modelar problemas ecológicos (RODRIGUES; TOMÉ, 2008).

O objetivo deste artigo não é apresentar uma revisão das várias abordagens teóricas em dinâmica de populações, mas sim apresentar as características básicas da dinâmica (estabilidade) de modelos populacionais do tipo Lotka-Volterra.

O artigo é apresentado na seguinte forma: na seção 2 descreve-se o modelo de Lotka-Volterra, seus pontos de equilíbrio e sua dinâmica/estabilidade em torno desses pontos críticos. Na seção 3 mostra-se que a introdução de um termo de saturação de presas, do tipo Verhulst, conduz a uma dinâmica mais rica, inclusive permitindo a modelagem de sistemas com extinção. O equilíbrio e a estabilidade desses modelos são analisados. Na seção 4 estuda-se o modelo de Monod-Haldane, um modelo do tipo Lotka-Volterra com termo de resposta funcional. Na seção 5 discutem-se os resultados deste trabalho.

## 2 MODELO DE LOTKA –VOLTERRA

Neste trabalho consideram-se modelos de duas espécies, uma espécie presa e a outra predadora. Tais sistemas não descrevem, no caso geral, as complexas relações observadas na Natureza, no entanto o estudo de modelos simples (de duas espécies interagindo) é um passo importante para a compreensão de fenômenos mais complexos. A modelagem matemática de sistemas com mais de duas espécies interagentes é simples, contudo a interpretação da dinâmica das populações torna-se complexa. Portanto, almejando um texto didático, restringe-se este estudo à dinâmica de duas populações.

Sejam as populações da presa e do predador, respectivamente, denotadas por $x(t)$ e $y(t)$, no instante $t$. Ao modelar matematicamente a interação das espécies, considera-se que na ausência do predador, $y(t)=0$, a população de presas aumentará, sem nenhum tipo de obstáculo, a uma taxa proporcional à população atual, ou seja, com um termo da forma $[ax(t)]$, onde $a$ é uma constante positiva. Por outro lado, considera-se que a carência de presas, $x(t)=0$, acarretará a extinção da população de predadores, devido à falta de alimento, situação descrita por um termo da forma $[-cy(t)]$, onde $c$ é uma constante positiva. Considera-se também que o número de encontros entre as duas espécies é proporcional ao produto das populações de cada espécie, ou seja, $x(t)y(t)$. Estes encontros tendem a promover o crescimento da população de predadores e a inibir o crescimento da população de presas. Assim, a taxa de crescimento da população de predadores, $dy/dt$, é aumentada por um termo da forma $[+\gamma\, x(t)y(t)]$, enquanto a taxa de crescimento da população de presas, $dx/dt$, é diminuída por um termo da forma $[-\alpha\, x(t)y(t)]$, onde $\alpha$ e $\gamma$ são constantes positivas. Em consequência dessa modelagem matemática, somos levados às equações de Lotka-Volterra:

$$\frac{dx}{dt} = ax - \alpha\, xy = x(a - \alpha\, y)$$

$$\frac{dy}{dt} = -cy + \gamma\, xy = y(-c + \gamma\, x) \tag{1}$$

Em (1), $a$ é a taxa de crescimento efetiva da população das presas na ausência de predadores, $c$ é a taxa de mortalidade da população de predadores na ausência de presas, $\alpha$ é a taxa de decréscimo da população de presas devido aos encontros com predadores e $\gamma$ é a taxa de crescimento da população dos predadores devido à predação.

Na literatura o modelo de Lotka-Volterra é adaptado a várias situações. Por exemplo, podem-se modelar matematicamente os efeitos da variação da temperatura do ambiente, que afetam a taxa de crescimento das populações, por meio da variação dos parâmetros $a, c, \alpha$ e $\gamma$ ao longo de um ciclo (dia, ano, etc..). Assim, em tais modelos de dependência sazonal, as taxas de variações das populações tornam-se dependentes das variações da temperatura.

Considere o sistema predador-presa descrito pelas equações de Lotka-Volterra (1), com a seguinte parametrização, $a = 1.0$, $c = 0.75$, $\alpha = 0.5$ e $\gamma = 0.5$, ou seja,

$$\frac{dx}{dt} = x(1.0 - 0.5\,y) \quad , \quad \frac{dy}{dt} = y(-0.75 + 0.5x) \; . \qquad (2)$$

Os pontos de equilíbrio de (2) são obtidos ao se tomar $dx/dt = dy/dt = 0$, ou seja,

$$x(1.0 - 0.5y) = 0 \quad , \quad y(-0.75 + 0.5x) = 0 \qquad (3)$$

As soluções de (3) fornecem os pontos de equilíbrio de (2), ou seja, os pontos (0, 0) e (1.5, 2.0). Em seguida será analisada a dinâmica do sistema (2) em torno de cada um desses pontos de equilíbrio, o que permitirá conclusões a respeito da estabilidade do sistema predador-presa descrito pelo sistema (2).

- <u>Dinâmica na vizinhança de (0, 0)</u>

O objetivo aqui é obter o comportamento (dinâmica) do sistema (2) em torno do ponto de equilíbrio (0, 0). Como o sistema (2) é quase linear (BOYCE; DIPRIMA, 2006) na vizinhança de (0, 0), ou seja,

$$\lim_{\substack{x\to 0\\ y\to 0}} \left[ \frac{\text{parte não linear de (2a)}}{\text{parte linear de (2a)}} \right] = \lim_{\substack{x\to 0\\ y\to 0}} \frac{-0.5\,xy}{1.0\,x} = 0$$

$$\lim_{\substack{x\to 0\\ y\to 0}} \left[ \frac{\text{parte não linear de (2b)}}{\text{parte linear de (2b)}} \right] = \lim_{\substack{x\to 0\\ y\to 0}} \left[ \frac{0.5\,xy}{-0.75\,y} \right] = 0 \, ,$$

pode-se desprezar nesse limite os termos não lineares nas equações (2), obtendo-se o sistema linear equivalente a seguir

$$\frac{d}{dt}\begin{pmatrix} x \\ y \end{pmatrix} = \begin{pmatrix} 1.0 & 0 \\ 0 & -0.75 \end{pmatrix} \begin{pmatrix} x \\ y \end{pmatrix} \; . \qquad (4)$$

O sistema linearizado (4) apresenta os autovalores $r_{1,2}$ e autovetores $\xi_{1,2}$ abaixo

$$r_1 = 1 \quad , \quad \xi_1 = \begin{pmatrix} 1 \\ 0 \end{pmatrix} \qquad \text{e} \qquad r_2 = -0.75 \quad , \quad \xi_2 = \begin{pmatrix} 0 \\ 1 \end{pmatrix} \quad , \tag{5}$$

de modo que sua solução geral é (BOYCE; DIPRIMA, 2006)

$$\begin{pmatrix} x \\ y \end{pmatrix} = c_1 \begin{pmatrix} 1 \\ 0 \end{pmatrix} e^{t} + c_2 \begin{pmatrix} 0 \\ 1 \end{pmatrix} e^{-0.75t} \qquad \text{ou} \qquad \begin{matrix} x(t) = c_1 \, e^{t} \\ y(t) = c_2 \, e^{-0.75t} \, . \end{matrix} \tag{6}$$

Note que a dinâmica das populações na vizinhança do ponto de equilíbrio de (0, 0) pode ser descrita tanto pelo sistema não linear (2) como pelo sistema linear (4). Note que o sistema linear (4) apresenta autovalores reais com sinais trocados, de modo que a origem é um ponto do tipo sela para os sistemas (2) e (4). Por este motivo, o ponto (0,0) é um ponto de equilíbrio instável.

- <u>Dinâmica na vizinhança de (1.5, 2.0)</u>

Deve-se escrever o sistema (2) em torno do ponto de equilíbrio (1.5, 2.0). Fazendo a mudança de variáveis $x = 1.5 + u$ *e* $y = 2.0 + v$ (translação) nas equações (2), obtêm-se

$$\frac{du}{dt} = v(-0.75 - 0.5u) \qquad , \qquad \frac{dv}{dt} = u(1.0 + 0.5v) \, . \tag{7}$$

Novamente verifica-se que o sistema (7) é quase linear em torno desse ponto. Assim, ignorando os termos não lineares nas variáveis $u$ e $v$, tem-se o sistema linear

$$\frac{d}{dt}\begin{pmatrix} u \\ v \end{pmatrix} = \begin{pmatrix} 0 & -0.75 \\ 1.0 & 0 \end{pmatrix} \begin{pmatrix} u \\ v \end{pmatrix} \, . \tag{8}$$

Os autovalores $r_{1,2}$ e autovetores $\xi_{1,2}$ do sistema linearizado acima são

$$r_1 = \left( \sqrt{3}\big/2 \right) i \quad , \quad \xi_1 = \begin{pmatrix} 1 \\ -2\left( \sqrt{3}\big/3 \right) i \end{pmatrix} \quad \text{e} \quad r_2 = -\left( \sqrt{3}\big/2 \right) i \quad , \quad \xi_2 = \begin{pmatrix} 1 \\ 2\left( \sqrt{3}\big/3 \right) i \end{pmatrix} \quad , \tag{9}$$

onde *i* é a unidade imaginária. Portanto, o ponto (1.5, 2.0) é do tipo centro para o sistema linear (8) e, consequentemente, um ponto de equilíbrio estável. Resolvendo explicitamente o sistema não-linear (2) é possível mostrar que o seu gráfico é uma curva fechada em torno do ponto de equilíbrio (1.5, 2.0). Logo, esse ponto também é do tipo centro para o sistema não linear (2), de modo que as populações de predadores e presas exibem uma variação cíclica, como pode ser conferido na Figura 1, para três condições iniciais.

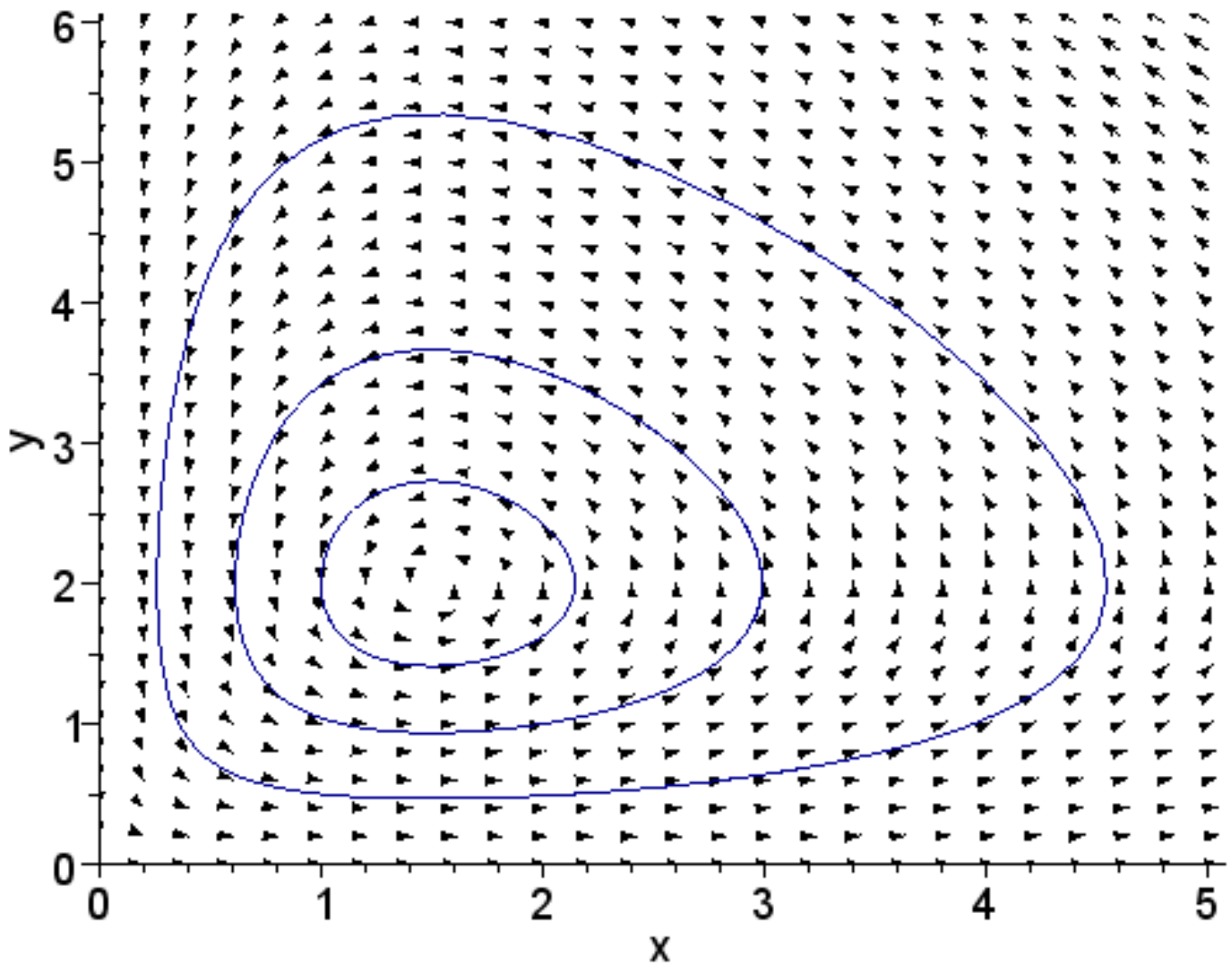

Figura 1: Representação da variação cíclica das populações descritas pelo modelo de Lotka-Volterra (2) para as condições iniciais (1.0, 0.5), (2.0, 1.0) e (1.0, 2.0).

A Figura 2 exibe o comportamento das populações de presas, $x(t)$, e de predadores, $y(t)$, em função do tempo $t$, para a condição inicial $(x_0, y_0) = (1.0, 0.5)$. Veja que inicialmente a população de presas (curva azul) cresce e, com a maior disponibilidade de alimento, a população de predadores (curva verde) também cresce. Em seguida, devido a maior predação, a população de presas diminui, e a população de predadores atinge o máximo suportado pelo sistema e, posteriormente, por falta de alimento, começa a diminuir. Enfim, com a diminuição da população de predadores, a população de presas volta a crescer e ciclo repete-se. Dessa forma o sistema oscila indefinidamente.

As figuras 1 e 2 mostram as oscilações típicas de populações descritas pelo modelo de Lotka-Volterra (1), independentemente dos valores escolhidos para os parâmetros $a, c, \alpha \text{ e } \gamma$ e das condições iniciais $(x_0, y_0)$. Um dos sistemas mais estudados por esta modelagem é o das populações de lebres e linces do Canadá. Registros realizados desde 1845 mostram uma clara variação periódica com períodos de 9 a 10 anos e com defasagem de 1 a 2 anos nos picos das populações de lebres e linces (ODUM; BARRET, 2007).

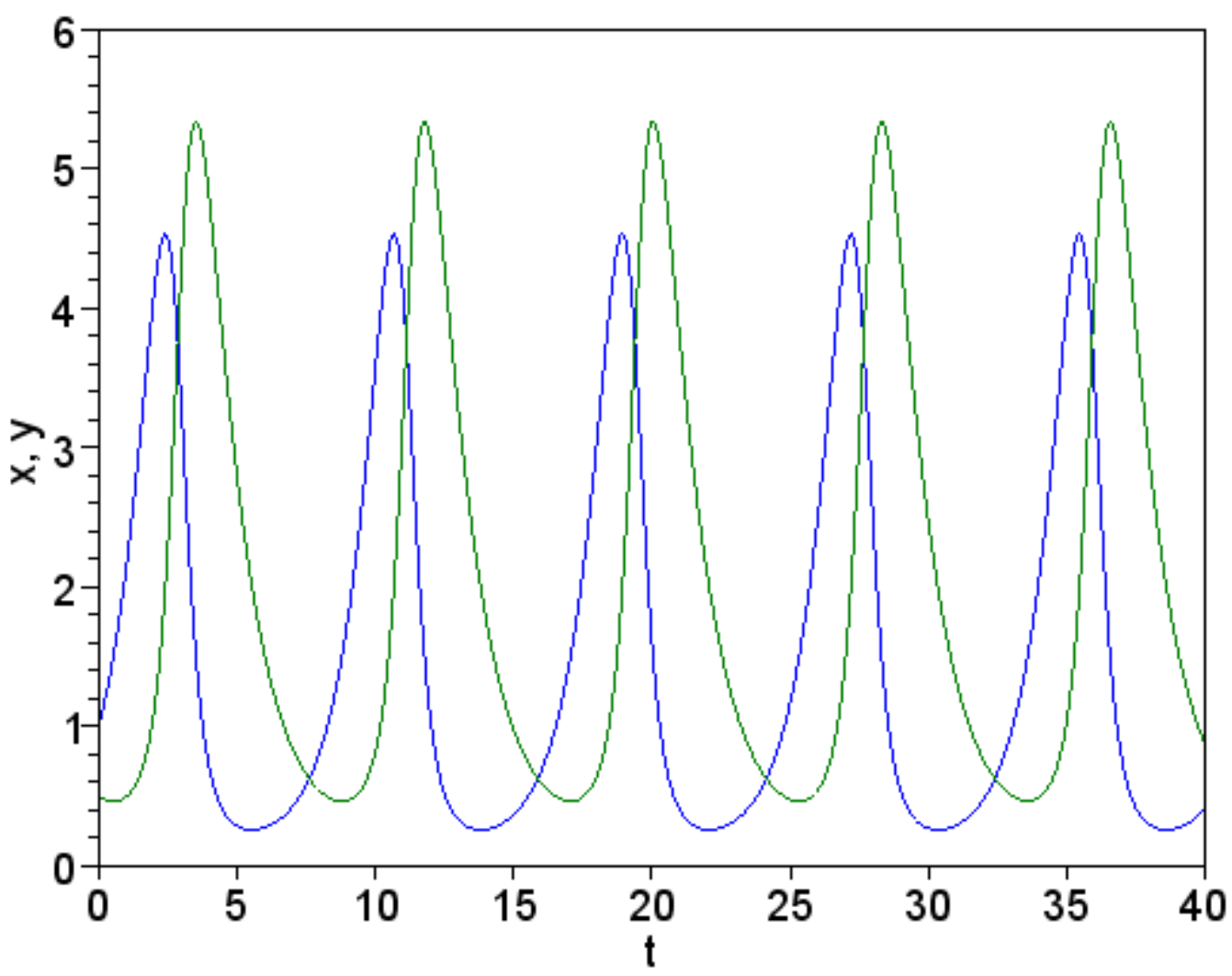

Figura 2: Variações das populações de presa (curva azul) e de predador (curva verde) em relação ao tempo *t* para a condição inicial (1.0, 0.5).

Por outro lado, muitos sistemas do tipo predador-pressa observados na natureza apresentam outros tipos de comportamentos, além do comportamento oscilatório. Observam-se sistemas que evoluem para populações assintoticamente estáveis ou sistemas em que populações se extinguem. Na próxima seção vamos apresentar equações do tipo Lotka-Volterra que descrevem tais situações.

## 3 MODELOS DE LOTKA–VOLTERRA APRIMORADOS

O modelo de Lotka-Volterra (1) não descreve sistemas biológicos do tipo predador-presa que evoluem para uma solução assintoticamente estável ou que apresentam extinção de populações. Para descrever tais sistemas biológicos verifica-se que a inclusão de um termo de saturação na população de presas, um termo logístico, permite o amortecimento das oscilações de Lotka-Volterra. Nas próximas subseções estudaremos a estabilidade de tais modelos.

### 3.1 Modelos de Lotka–Volterra com amortecimento

Considere o sistema de Lotka-Volterra (2) com um termo de saturação na população de presas $x(t)$, termo do tipo $\left(-k\,x^2\right)$ com *k* uma constante positiva,

$$\frac{dx}{dt} = x\left(1.0 - 0.5\,y\right) - 0.5\,x^2 \quad , \quad \frac{dy}{dt} = y\left(-0.75 + 0.5\,x\right) \ . \qquad (14)$$

Os pontos de equilíbrio de (14) são dados pelo sistema

$$x(1.0 \text{ - } 0.5y \text{ - } 0.5x) = 0 \quad , \quad y(\text{-}0.75 + 0.5x) = 0 \ . \qquad (15)$$

Resolvendo o sistema (15) têm-se três pontos de equilíbrio: (0, 0), (2.0, 0) e (1.5, 0.5). Esses pontos são soluções de equilíbrio do sistema (14).

A seguir analisar-se-á a dinâmica do sistema (14) em torno de cada um desses pontos. Através do procedimento já realizado na seção anterior, verifica-se que o sistema (14) tem comportamento quase linear em torno dos três pontos de equilíbrio obtidos, de modo que se pode estudar o sistema (14) por meio do sistema linear equivalente (BOYCE; DIPRIMA, 2006).

- Dinâmica na vizinhança de (0, 0)

Ignorando os termos não lineares em (14), segue o sistema linear

$$\frac{d}{dt}\begin{pmatrix} x \\ y \end{pmatrix} = \begin{pmatrix} 1.0 & 0 \\ 0 & -0.75 \end{pmatrix}\begin{pmatrix} x \\ y \end{pmatrix} , \qquad (16)$$

já analisado na seção precedente. A solução geral de (16) é dada em (6), de modo que a origem continua sendo um ponto de sela para os sistemas (14) e (16), ou seja, um ponto de estabilidade instável.

- Dinâmica na vizinhança de (2.0, 0)

Note que este ponto de equilíbrio corresponde à situação de ausência de predadores, $y(t) = 0$ para todo $t$. Neste caso, qualquer que seja a população, não nula, inicial de presas, ela evoluirá assintoticamente para a população de saturação de presas, $x(t \to \infty) = 2$.

- Dinâmica na vizinhança de (1.5, 0.5)

Deve-se linearizar o sistema (14) em torno deste ponto de equilíbrio. Seja a translação dada pela mudança de variáveis

$$x = 1.5 + u \quad , \quad y = 0.5 + v \qquad (17)$$

nas equações (14). Ignorando os termos não lineares em $u$ e $v$ tem-se o sistema linear equivalente

$$\frac{d}{dt}\begin{pmatrix} u \\ v \end{pmatrix} = \begin{pmatrix} -0.75 & -0.75 \\ 0.25 & 0 \end{pmatrix}\begin{pmatrix} u \\ v \end{pmatrix} \quad , \tag{18}$$

que apresenta os seguintes autovalores complexos

$$r_1 = -\frac{3}{8} + \frac{\sqrt{3}}{8}i \quad , \quad r_2 = -\frac{3}{8} - \frac{\sqrt{3}}{8}i \quad . \tag{19}$$

Como os autovalores são imaginários, com parte real negativa, tem-se que o ponto de equilíbrio (1.5, 0.5) é um ponto espiral assintoticamente estável.

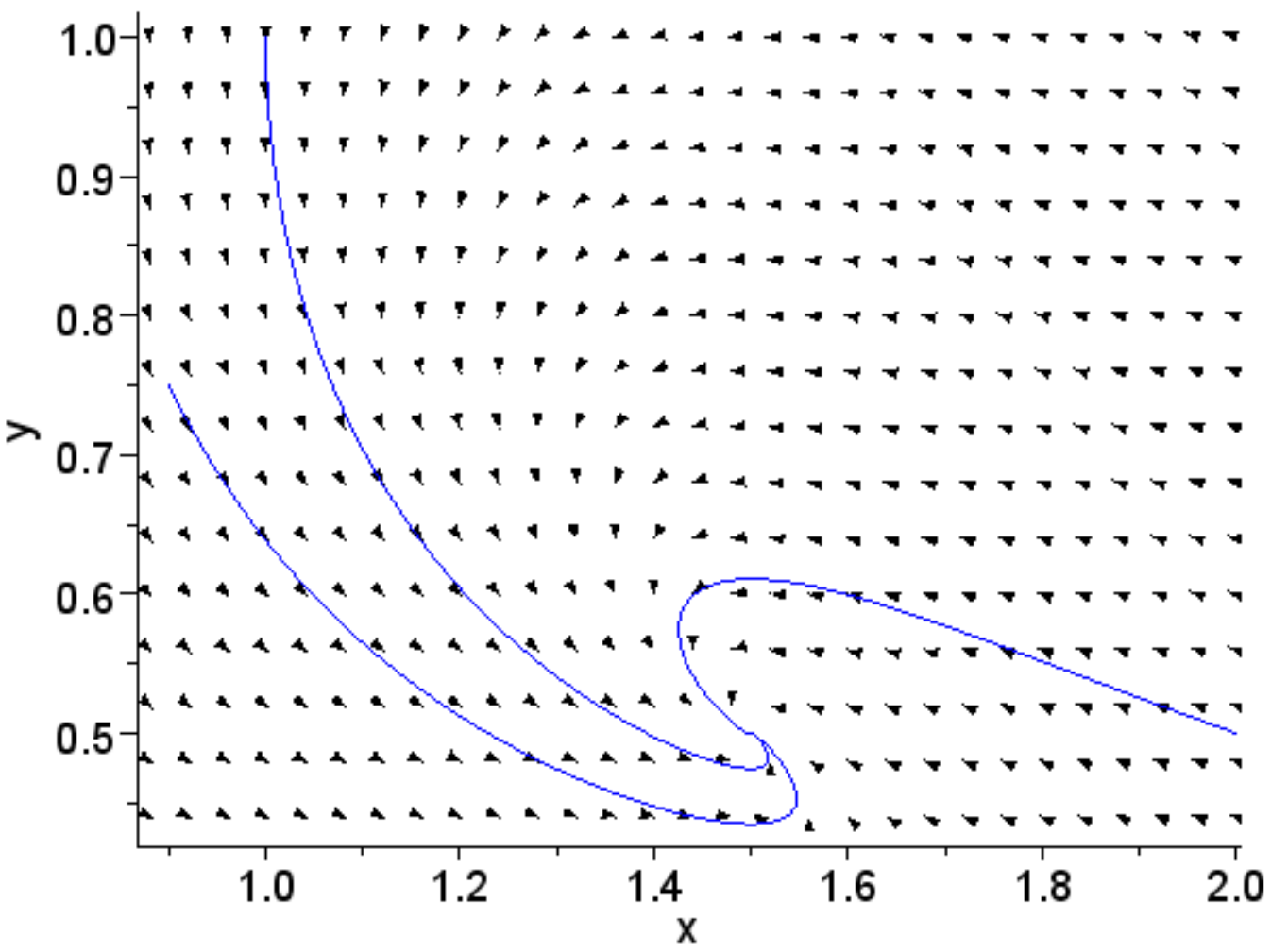

Figura 3: Campo de direções do sistema (14). As curvas são as soluções particulares para as condições iniciais: (0.9, 0.75), (1.0, 1.0) e (2.0, 0.5).

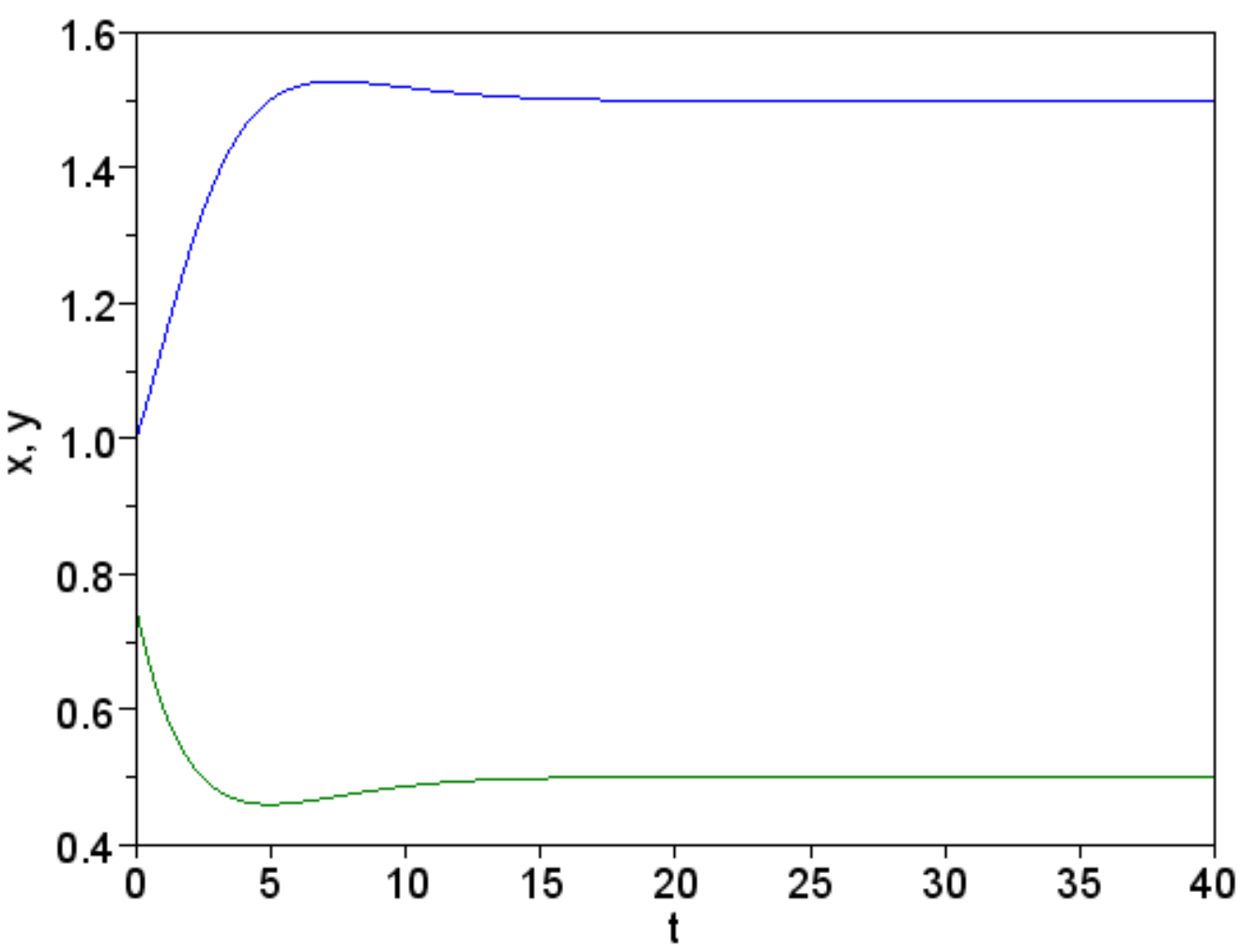

Figura 4: Variações amortecidas das populações de presa (curva azul) e de predador (curva verde) do modelo de Lotka-Volterra (14) para a condição inicial (1.0, 0.75).

A figura 3 exibe o campo de direções do sistema (14) para três condições iniciais. Observe que as soluções, independentemente das condições iniciais, evoluem assintoticamnete para a solução estacionária $(x_\infty, y_\infty) = (1.5, 0.5)$. A figura 4 exibe a dependência das populações $x$ e $y$ com o tempo $t$. Note que para $t > 20$ as populações atingem uma situação estacionária.

### 3.2 Modelos de Lotka–Volterra com extinção

Nesta seção pretende-se modelar o fenômeno de extinção em sistemas predador-presa. Mostrar-se-á que tal comportamento também pode ser descrito por um sistema de Lotka-Volterra com termo de saturação na população de presas. A diferença fundamental entre o sistema desta seção e aquele apresentado na seção anterior é o valor dado ao parâmetro $c$. Nesta aplicação tomar-se-á $c = 1.5$, assim aumentando-se a dependência da sobrevivência da população de predador com respeito à ausência de presas, ou ainda, com a diminuição desta última população.

Considere o sistema (14) com $c = 1.5$, ou seja,

$$\frac{dx}{dt} = x(1.0 - 0.5y) - 0.5x^2 \quad , \quad \frac{dy}{dt} = y(-1.5 + 0.5x) \ . \tag{20}$$

Resolvendo o sistema de equações

$$x(1.0 - 0.5y - 0.5x) = 0 \quad , \quad y(-1.5 + 0.5x) = 0 \tag{21}$$

obtêm-se três pontos de equilíbrio do sistema (20), ou seja, (0, 0), (2.0, 0) e (3.0,-1.0). No entanto, o ponto (3.0,-1.0) corresponde a uma solução estacionária irreal, pois não existe população negativa. Novamente, através do procedimento já descrito na seção anterior, verifica-se que o sistema (20) tem comportamento quase linear em torno dos dois pontos de equilíbrio considerados, de modo que se pode estudar o sistema (20) por meio do sistema linear equivalente.

- Dinâmica na vizinhança de (0, 0)

Desprezando os termos não lineares nas equações (20), obtém-se o sistema linear equivalente

$$\frac{d}{dt}\begin{pmatrix} x \\ y \end{pmatrix} = \begin{pmatrix} 1.0 & 0 \\ 0 & -1.5 \end{pmatrix}\begin{pmatrix} x \\ y \end{pmatrix} \ . \tag{22}$$

com autovalores da matriz de coeficientes dados por

$$r_1 = 1 \quad , \quad r_2 = -1.5 \quad . \qquad (23)$$

Novamente, como os autovalores são reais de sinais trocados, segue que a origem é um ponto de sela para os sistemas (20) e (22) e, portanto, um ponto de estabilidade instável.

- Dinâmica na vizinhança de (2.0, 0)

Considere a seguinte mudança de variáveis

$$x = 2.0 + u \quad , \quad y = v \qquad (24)$$

nas equações (20). Ignorando os termos não lineares em $u$ e $v$, obtém-se

$$\frac{d}{dt}\begin{pmatrix} u \\ v \end{pmatrix} = \begin{pmatrix} -1.0 & -1.0 \\ 0 & -0.5 \end{pmatrix}\begin{pmatrix} u \\ v \end{pmatrix} \qquad (25)$$

com os seguintes autovalores

$$r_1 = -1.0 \quad \text{e} \quad r_2 = -0.5 \quad . \qquad (26)$$

Agora, como os autovalores são reais negativos, o ponto (2.0, 0) é um ponto assintoticamente estável do sistema linearizado (25) e do sistema (20). Assim, dadas condições iniciais, as trajetórias de (20) se aproximam do ponto de equilíbrio (2.0, 0), quando $t \to \infty$, de modo que o ponto estudado corresponde à ocorrência da extinção da população de predadores. A Figura 5 representa o campo de direções do sistema (20) para várias condições iniciais.

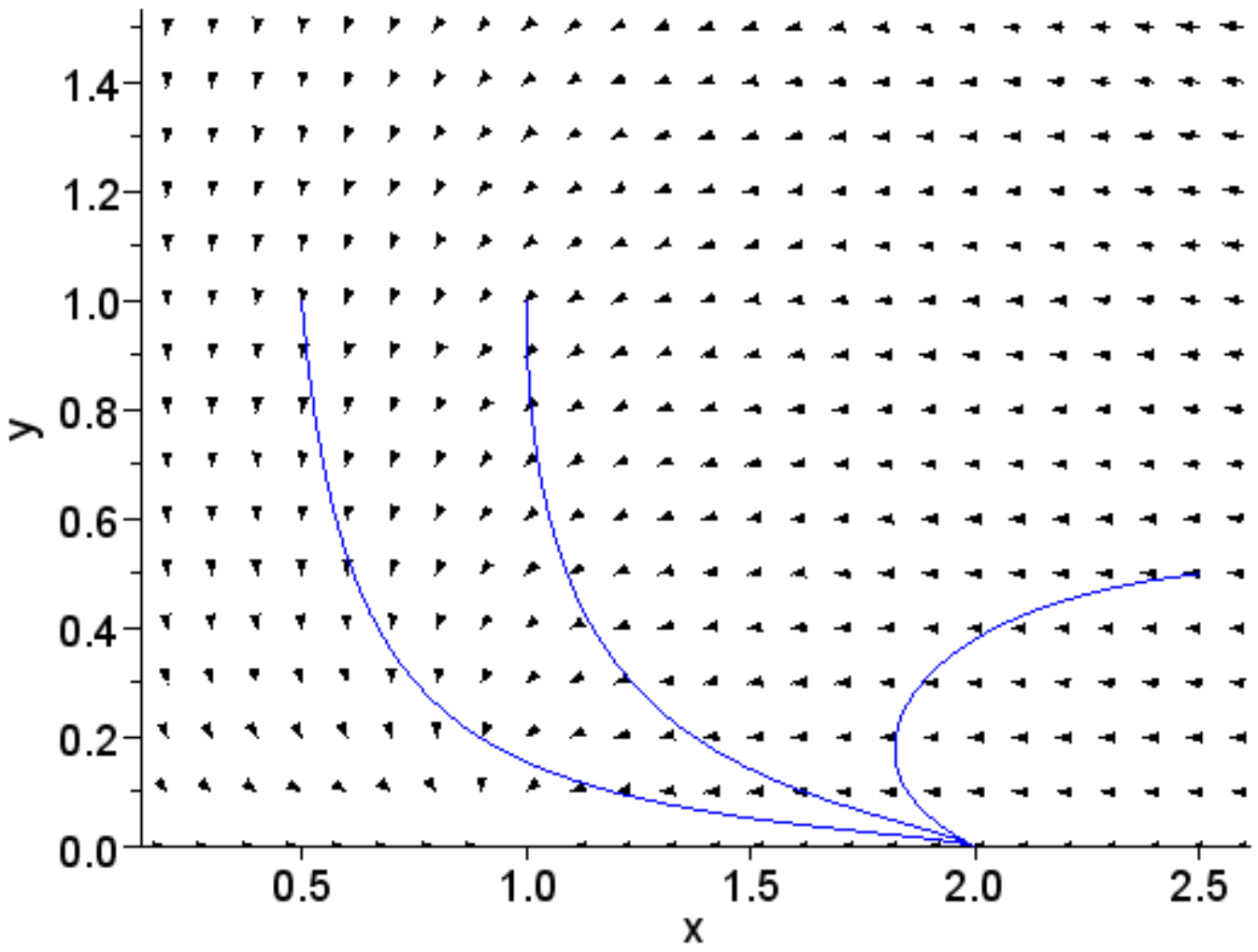

Figura 5: Campo de direções do sistema (20). As curvas são as soluções particulares (com extinção) para as condições iniciais: (0.5, 1.0), (1.0, 1.0) e (2.5, 0.5).

A Figura 6 mostra a dependência de $x$ e $y$, em $t$, para a condição inicial (1.0, 1.0). Note que, devido à alta taxa de mortalidade dos predadores, $c = 1.5$, a população de predadores diminui até a extinção, enquanto a população de presas atinge sua saturação.

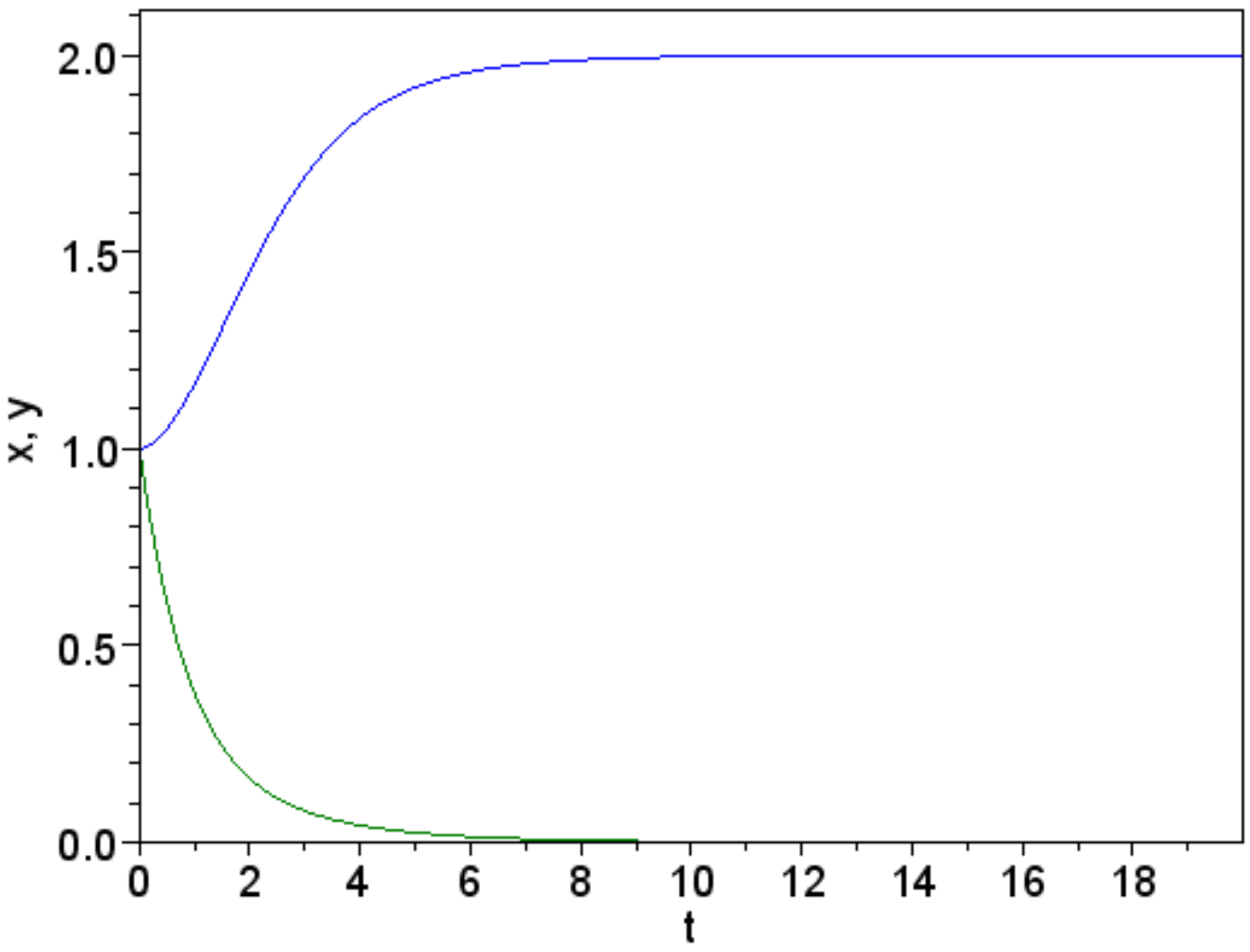

Figura 6: Populações de presa (curva azul) e de predadores (curva verde) descritas pelo modelo de Lotka-Volterra (20), em relação ao tempo *t*, para a condição inicial (1.0, 1.0).

## 4 MODELOS DE MONOD-HALDANE

Uma variação importante do modelo de Lotka-Volterra é o modelo de Monod-Haldane. Jacques Lucien Monod (1910-1976) foi um biólogo francês premiado com o Prêmio Nobel de Medicina em 1965 e um dos fundadores da Biologia Molecular. John Burdon Sanderson Haldane (1892-1964) foi um biólogo britânico e um dos fundadores da Genética Populacional.

Inicialmente, Holling (1963) propôs três diferentes tipos de respostas funcionais para sistemas predador-presa, todas monotonamente crescentes. Por outro lado se observa respostas não monotônicas em alguns sistemas predador-presa. Uma resposta funcional do tipo Monod-Haldane é adequada para sistemas que apresentam algum tipo de efeito inibitório na resposta dos predadores às mudanças na população de presas (ZHANG; LI, 2014). Tais respostas funcionais são atualmente também chamadas de respostas funcionais Holling do tipo IV. Uma revisão sobre estes modelos com resposta funcional é encontrada nas referências (HASTINGS, 1997; MAY, 2001).

Modelos de Monod-Haldane são da forma (BAEK; DO; SAITO, 2009)

$$\frac{dx}{dt} = ax\,(1 - \frac{x}{K}) \; - \; \frac{\alpha\, xy}{(1 + bx^2)}$$

$$\frac{dy}{dt} = y\,[\; \frac{\gamma\, x}{(1 + bx^2)} - c \;] \quad , \tag{28}$$

onde $a$, $c$, $b$, $K = a/k$, $\alpha$ e $\gamma$ são constantes positivas com $a$ a taxa de nascimento de presas, $c$ a taxa de mortalidade de predadores, $b$ o nível de interferência da população de presas na predação, *K* a taxa de saturação da população de presas e $\alpha$ e $\gamma$ as taxas de interação entre presas e predadores.

Considere a parametrização do sistema (14) em (28), com o nível de interferência da população de presas na predação dado por $b = 0.05$. Note que temos uma interferência baixa. Segue o seguinte sistema de Monod-Haldane

$$\frac{dx}{dt} = x(1 - 0.5x) - \frac{0.5\,xy}{(1 + 0.05x^2)}$$

$$\frac{dy}{dt} = y\left( \frac{0.5\,x}{(1 + 0.05x^2)} - 0.75 \right) . \tag{29}$$

Resolvendo as equações de equilíbrio de (29) e considerando que tais soluções (populações) devem ser reais e positivas, têm-se três pontos de equilíbrio para o sistema (29), ou seja, (0, 0), (2.0, 0) e (1.7225, 0.3186). Como nos casos anteriores, o sistema (29) tem comportamento quase linear em torno destes pontos. Verifica-se que os pontos (0, 0) e (2.0, 0) são pontos de estabilidade instável (autovalores reais com sinais trocados), enquanto o ponto de equilíbrio (1.7225, 0.3186) é assintoticamente estável (autovalores complexos com parte real negativa).

A Figura 7 representa o campo de direções do sistema (29) para várias condições iniciais. A Figura 8 mostra a dependência de $x$ e $y$, em $t$, para a condição inicial (1.4, 0.4). Note que o pequeno nível de interferência da população de pressas na predação, dado por $b = 0.05 \neq 0$, gera uma dinâmica diferente das dinâmicas observadas anteriormente. Nota-se, por exemplo, que o sistema predador-presa demora mais para atingir uma situação estacionária.

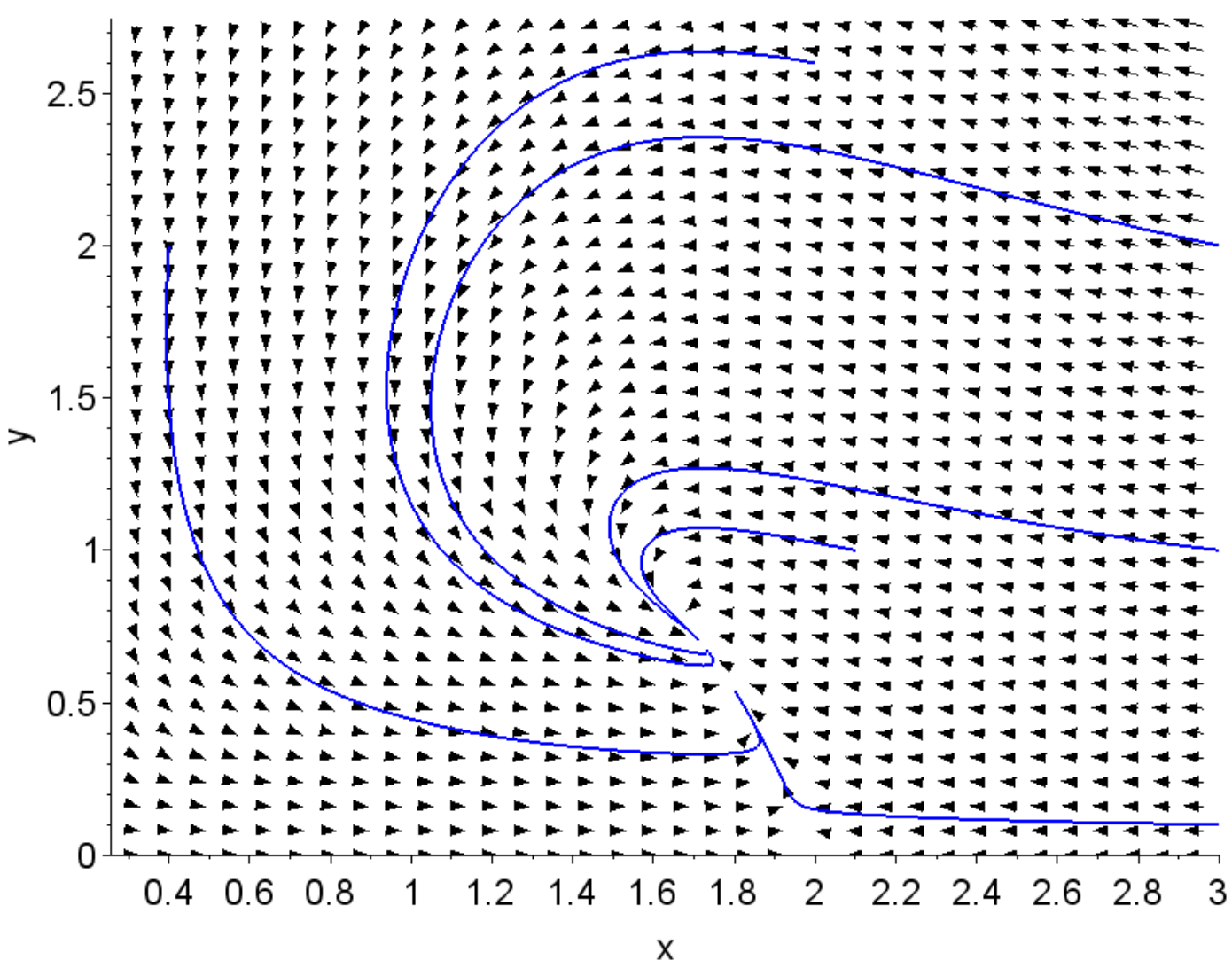

Figura 7: Campo de direções do sistema (29). As curvas cheias são as soluções particulares para as condições iniciais: (0.4, 2.0), (2.0, 2.5), (3.0, 2.0), (3.0, 1.0), (2.0, 1.0) e (3.0, 0.1).

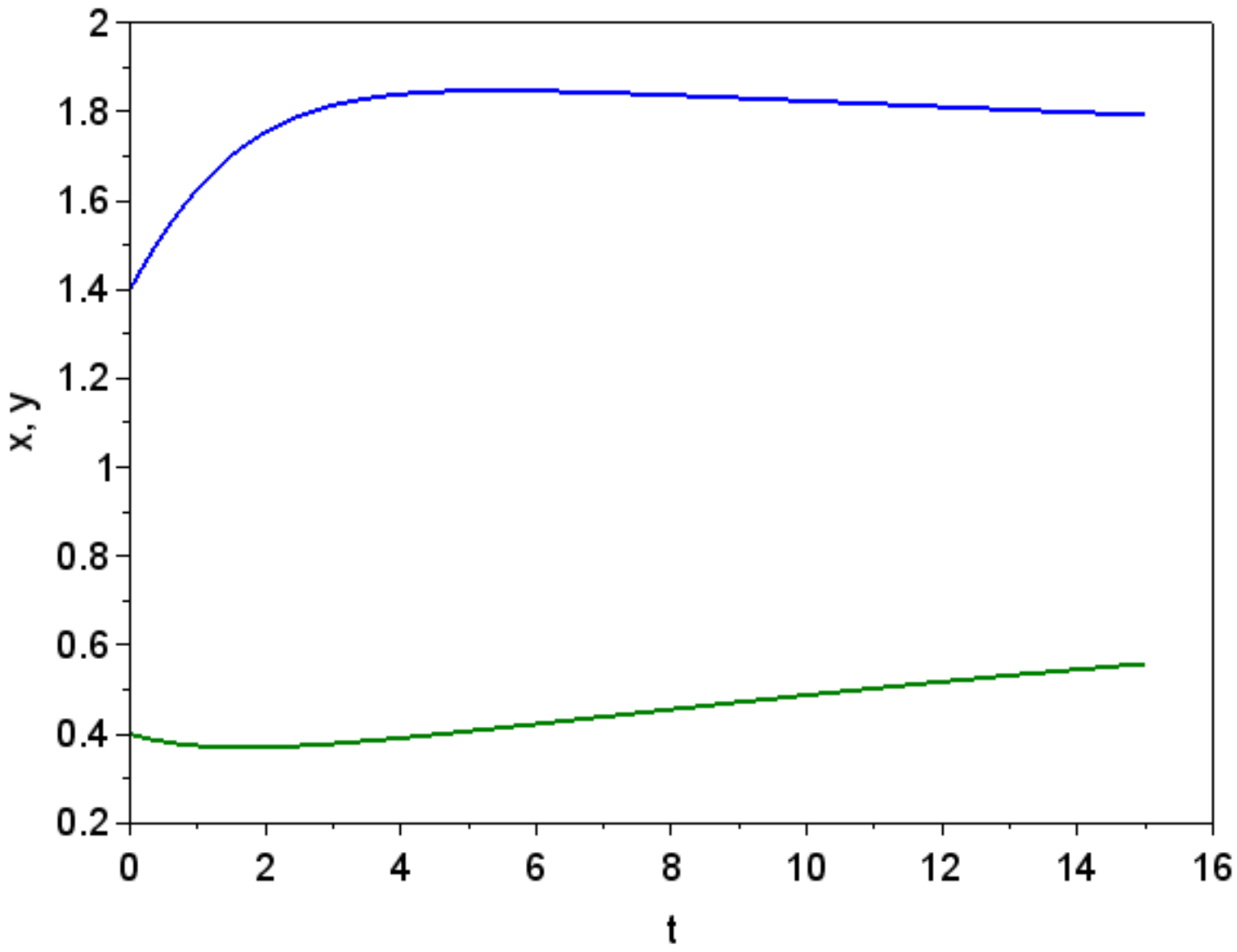

Figura 8: Populações de presa (curva azul) e de predadores (curva verde) descritas pelo modelo de Lotka-Volterra (29), em relação ao tempo *t*, para a condição inicial (1.4, 0.4).

## 5 CONCLUSÃO

Mostrou-se que sistemas de Lotka-Volterra podem descrever uma ampla variedade de sistemas do tipo predador-presa, desde sistemas em que as populações oscilam até sistema que atingem um equilíbrio após um intervalo de tempo (populações estacionárias). Verificou-se também que através desses modelos é possível prever se uma espécie está em perigo de extinção. Para esse estudo é importante conhecer a taxa de mortalidade da espécie predadora (parâmetro $c$), a taxa de natalidade da espécie presa (parâmetro $a$) e a taxa de encontro das espécies (parâmetro $\gamma$). Uma razão desfavorável desses fatores pode levar uma população à extinção.

Estudou-se também o efeito da resposta da população dos predadores às mudanças na população de presas, efeito chamado de resposta funcional. Dentre os modelos de resposta funcional foi dada maior atenção aos chamados modelos de Monod-Haldane. Em tais modelos, por meio de um parâmetro que descreve a interferência da população de presas na predação (parâmetro $b$), podem-se descrever vários fenômenos, inclusive a dificuldade que determinados sistema predador-presa têm para atingir situações estacionárias, necessitando de grandes intervalos de tempo.

Enfim, salienta-se que o estudo de sistemas com $n$ espécies, algumas predadoras outras presas, pode ser realizado por meio de um sistema de $n$ equações diferencias acopladas. Tais sistemas apresentam uma dinâmica complexa, geralmente caótica.

## REFERÊNCIAS